\input{aipcheck.tex}

\documentclass[final]{aipproc}

\layoutstyle{6x9}
  
\usepackage{epsfig}
\usepackage{here}
\usepackage{graphics}

\newcommand{\reaktion}{\mbox{$pp\rightarrow\,ppK^+K^-$}}

\begin{document}

\title{Open strangeness threshold production at COSY-11}


\newcommand{\ikpjuel}{IKP, Forschungszentrum J\"{u}lich, D-52425 J\"{u}lich, Germany}
\author{P.~Winter, for the COSY-11 collaboration}{address={\ikpjuel}}

\copyrightyear{2001}

\begin{abstract}
The open strangeness production near threshold is investigated at the internal experiment COSY-11 in different reaction channels. Recently, the main focus has been to extend measurements of the hyperon production $pp\to pK^+\Lambda/\Sigma^0$ in different isospin channels as well as the associated strangeness production in $pp\to ppK^+K^-$. The experimental technique is based on the reconstruction of the four momentum for all positively charged ejectiles. Neutrons are detected in addition using a neutral particle detector. The unregistered hyperon or meson is identified by means of the missing mass technique. The present status of the analysis for both reaction channels is presented. In case of the reaction \reaktion, very preliminary cross sections are shown.
\end{abstract}

\date{\today}

\maketitle

\section{Experimental technique}
\begin{figure}[h]
\epsfig{file=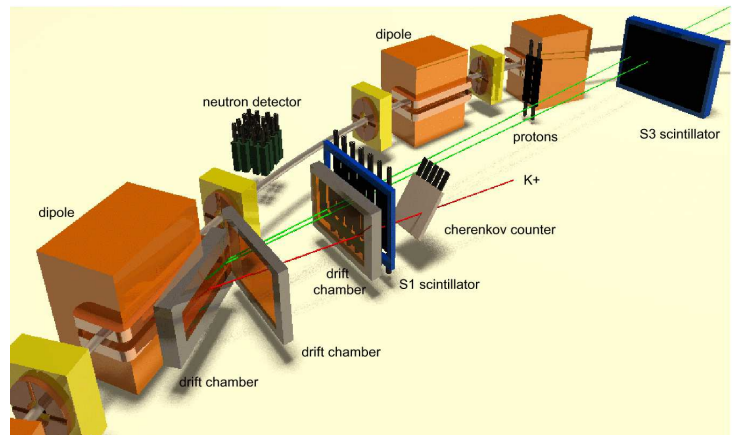,width=0.65\textwidth}
\caption{\label{cosy11} Schematic view of the COSY-11 setup for an exemplary event in \reaktion, where the kaon decays before reaching the stop scintillator S3. The not shown cluster target is located in front of the left dipole magnet.}
\end{figure}

The magnetic spectrometer COSY-11 \cite{brauksiepe:96} at the COoler SYnchrotron COSY \cite{maier:97nim} is shown in its principal layout in figure \ref{cosy11}.
The reaction takes place in a cluster target mounted in front of one of the ring dipoles and is operated with hydrogen or deuterium. The positively charged ejectiles are bent to the interior of the ring and their four momenta are reconstructed using the information of a set of drift chambers and a subsequent time of flight measurement. The $K^-$ for the \reaktion reaction is then identified via the missing mass method. In case of $pp\to nK^+\Sigma^+$, additionally the neutron is detected in a lead-scintillator arrangement and the unregistered $\Sigma^+$ is again deduced from the missing mass technique. Details on the explicit experimental technique and the analysis can be found elsewhere \cite{brauksiepe:96,moskal:01}.

\section{Hyperon production}
Close-to-threshold production data \cite{balewski:98-2,sewerin:99,kowina:04} in $pp\to pK^+\Lambda/\Sigma^0$ revealed a cross section ratio \mbox{$R=\sigma_{tot}(\Lambda)/\sigma_{tot}(\Sigma^0)\approx28$} exceeding that at high energies by an order of magnitude. Different models \cite{gasparian:99,shyam:04,sibirtsev:97} within the framework of one-boson exchange do not reproduce the energy dependence too well. In order to have further constraints to the theoretical descriptions, the COSY-11 collaboration collected new data in the different isospin channel $pp\to nK^+\Sigma^+$ at $Q=$\,13 and 60\,MeV.\\
\begin{figure}[H]
\epsfig{file=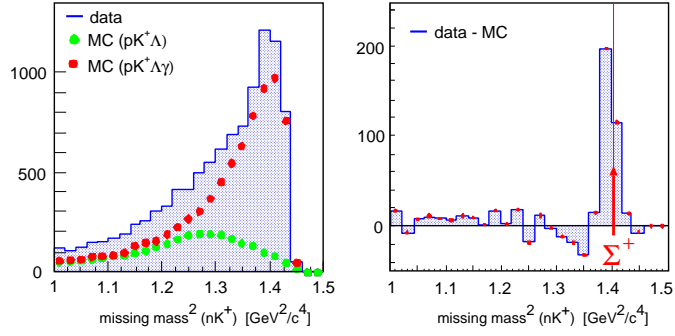,width=0.6\textwidth}
\caption{Left: Squared missing mass spectrum for the $nK^+$-system. Monte Carlo simulations for the main background channels $pp\to pK^+\Lambda$ and $pp\to pK^+\Lambda\gamma$ are shown by the dotted curves. Right: Subtraction of the background Monte Carlo spectra from the data.\label{missmass-nK}}
\end{figure}
For now, a missing mass spectrum for the lower excess energy with respect to the $nK^+$-system was elaborated \cite{rozek:04} and is presented in figure \ref{missmass-nK} left side. The dominant background channels were identified to be $pp\to pK^+\Lambda$ and $pp\to pK^+\Lambda\gamma$. This background was subtracted from the data (right plot in fig. \ref{missmass-nK}). Therefore, the known cross section for $pp\to pK^+\Lambda$ was incorporated and then the $pp\to pK^+\Lambda\gamma$ channel was fitted  such that the experimental missing mass spectrum below $1.3\,$GeV$^2$/c$^4$ was reproduced. Although a peak around the $\Sigma^+$ mass is seen, a lot more studies have to be performed before conclusive results on the cross section will be drawn.

\section{Elementary kaon production} 
\begin{figure}[h]
\rotatebox{-90}{\epsfig{file=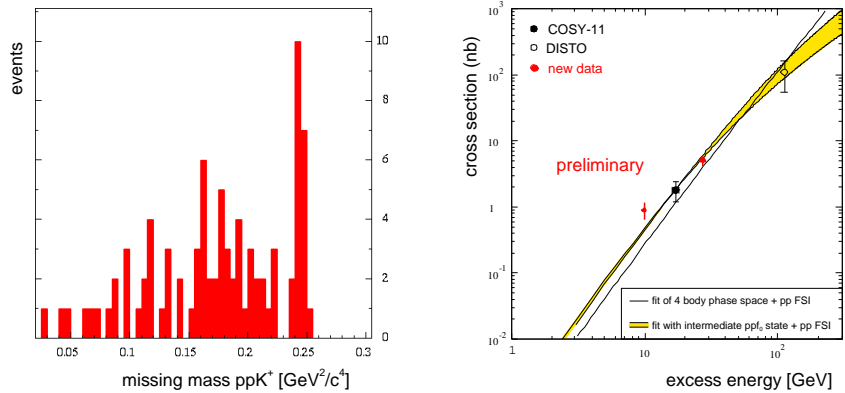,scale=0.5}}
\caption{\label{missmass}Left: Missing mass of the $ppK^+$-system with an additional demand for a hit in the dipole scintillator. Right: Excitation function of the total cross section for \reaktion. New data at $Q=10$ and $28\,$MeV are preliminary and include only statistical errors so far.}
\end{figure}

Exclusive data on the $pp\to ppK^+K^-$ reaction have been taken at COSY-11 at several excess energies motivated by the ongoing discussion about the nature of the scalar resonances $a_0$ and $f_0$. Besides the interpretation as a $q\bar{q}$ state \cite{morgan:93}, these resonances are also proposed to be $qq\bar{q}\bar{q}$ states \cite{jaffe:77}, $K\bar{K}$ molecules \cite{weinstein:90,lohse:90}, a hybrid $q\bar{q}$/meson-meson system \cite{beveren:86} or even quarkless gluonic hadrons \cite{jaffe:75}. In addition, final state interaction effects occur -- if existent -- strongly at low excess energies. Therefore, the elementary production process is a good tool to learn more about a possible $pK^-$ FSI or additional degrees of freedom in this four body final state.\\
A first total cross section $\sigma=1.80\pm0.27^{+0.28}_{-0.35}$\,nb for the excess energy $Q=17\,$MeV was determined in a former measurment at COSY-11 \cite{quentmeier:01-2} while further data at two other $Q$-values of 10 and 28\,MeV were taken. After the identification of two protons and a $K^+$, a clear signal in the missing mass spectrum at the kaon mass is observed for both energies (as an example at $Q=28\,$MeV see figure \ref{missmass}). Here, a hit in a scintillator mounted inside the dipole gap, where the $K^-$ is going to, was required.
There is no doubt that the background will be described by the hyperon channels $pp\to pK^+\Lambda(1405) / \Sigma(1385)$ and some misidentified pions like it was shown in the case of the data at $Q=17\,$MeV. Therefore, the full analysis (for details see \cite{quentmeier:01-2}) will most likely give also for the new data sets a background free missing mass spectrum. Within the near future, the final cross sections will be extracted. This however requires much deeper studies of the detection efficiency.


\bibliographystyle{aipproc}
\bibliography{abbrev,general}

\end{document}